\documentstyle[epsf,preprint,aps]{revtex}
\draft
\begin{document}
\title{\bf Optical Properties of BN in the cubic and in the layered 
hexagonal phases}
%\vspace*{1cm}
%\maketitle
\author{Giancarlo Cappellini and  Guido Satta}
\address{INFM Sezione di Cagliari and Dipartimento di Fisica, 
Universit\`a di Cagliari, Cittadella Universitaria, Strada Prov.le 
Monserrato-Sestu Km 0.700, I-09042 Monserrato (Ca), Italy}
%\maketitle
%\par
\author
{Maurizia Palummo and Giovanni Onida}
\address{INFM Sezione di Roma-2 and Dipartimento di Fisica, Universit\`a di 
Tor Vergata, via della Ricerca Scientifica 1, I-00133, Rome, Italy }
%\receipt{}
\maketitle
%\vspace*{1cm}
\begin{abstract}
\noindent
Linear optical functions of cubic and hexagonal BN have been 
studied within first principles DFT-LDA theory. Calculated 
energy-loss functions compare well with experiments and 
previous theoretical results both for h-BN and for c-BN. 
Discrepancies arise between theoretical results and experiments 
in the imaginary part of the dielectric function for c-BN. 
Possible explanation to this mismatch are proposed and 
evaluated; lattice constant variations, h-BN contamination in c-BN 
samples and self-energy effects. 
%
%On the other hand for h-BN our 
%results show a reasonable agreement with other theoretical 
%results and experiments for the imaginary part of the optical 
%functions. 
\end{abstract}
%\vspace*{1cm}
%\vspace*{1cm}
%\maketitle
%\begin{multicols}2
\narrowtext
\section{Introduction}
The properties of boron nitride (BN) have  motivated detailed theoretical and 
experimental studies since a long time \cite{kleinmann,doni,philiz}.
Many advanced technologies rely on Boron Nitride and on materials
based on it, due the wide spectrum of properties offered by its polymorphic
modifications, two graphite-like and two dense ones.
Boron nitride shares many of its properties, structures, 
processings and applications with carbon. 
Cubic boron nitride (also known as sphalerite boron nitride and 
abbreviated as Z-BN, c-BN or $\beta$-BN),with $sp^3$-hybridized
B-N bonds, has the diamond  crystal structures and a similar lattice 
constants. Its physical properties, such as extreme hardness, 
wide energy bandgap, low dielectric constant and high thermal 
conductivity, are also very near to those of diamond. 
These amazing properties of c-BN have many appealing 
applications in modern microeletronic devices, and make  it 
useful also as a protective coating material or in high-duty 
tools \cite{edgar}.
Hexagonal BN (h-BN or $\alpha$-BN), an $sp^2$-bounded layered compound and 
graphite also resemble each other in term of  
crystal structures, lattice constants, and physical properties 
such as strong anisotropy. 
%
%h-BN , a $sp^2$-bonded layered compound isostructural to graphite,
%exhibits strong anisotropic physical properties.
%
Due to its high termal stability h-BN is a widely used 
material in vacuum technology. It has been employed in microelectronic devices
\cite{pauli}, for x-ray lithography masks \cite{dana},
and as a wear-resistant lubrificant \cite{miyoshi}. 
The hexagonal phase is also the underlying structure of the BN nanotubes,
 which are systems of growing interest nowadays
\cite{blase1,rubio,blase2}. 
\par
On the other hand, significant differences exist between 
carbon and boron nitride, due mainly to differences in their chemical 
bonding. Infact h-BN and c-BN  mechanical strengths, 
thermal conductivities and Debye temperatures are lower than in their carbon 
counterparts. Moreover h-BN is electrically an insulator, while 
graphite is a conductive semi-metal. 
In contrast to diamond, which can be readly doped only as a p-type, c-BN  can 
be doped either n- or p-type. 
Moreover c-BN does not react with ferrous materials, even at high 
temperatures (T$\sim 1600 K$) \cite{vel},
and, last but not least, both c- and h-BN are more resistent 
to oxidation than their carbon counterparts \cite{edgar}.
\par
An important issue is the
experimental evidences for negative electron affinity (NEA)
in cubic boron nitride samples \cite{powers,pryor}.
The electron affinity of a semiconductor is the difference between 
the vacuum energy level and the conduction band minimum level. 
NEA occurs when the conduction band minimum lies above the vacuum 
energy level. Any electron promoted into the conduction band 
has then enough energy to escape into vacuum. 
One of the most striking application is in electron 
emission devices, to obtain the highest electron emission density at 
the least energy expenditure (cold cathode emitters) \cite{williams}.
Correct evaluation of the one electron transition energies and the 
position of the single particle states turn out to be fundamental
points relatively to this issue, and will be here addressed in detail.
\par
It is now generally agreed experimentally that bulk c-BN is the 
thermodynamically stable phase at ambient conditions, and that the 
less dense h-BN becomes stable at temperatures exceding $T \sim1200K$
\cite{eremets}.
Due to the large change in volume,
fragmentation and disordering between the two phases 
can develop. In combination with a rigid lattice, this 
lead to a large hysteresis of phase transitions. 
This means that phases created under high pressure 
high-temperature conditions can persist inside the sample
under standard condition \cite{eremets}.
Two other phases, the rhombohedral(r-BN) and the
wurtzite (w-BN) are stable only at very high pressure 
(P$\simeq$10GPa) \cite{solozhenko1} and will not be considered here.
Structural and excitation properties of wurtzite structure has
been studied elsewhere \cite{cappefiore}.
At ambient pressures an interplay between h-BN and c-BN domains only should be 
expected \cite{solozhenko2,kern}.
Since 1979 c-BN chemical vapor deposited films  have been realized
\cite{sokolowski}, but the production of pure c-BN thin films (either by 
chemical or physical vapor deposition) remains  a difficult 
task due to the formation, during the growth process, of unwilled h-BN 
domains \cite{kern}.
\par 
Extensive theoretical studies have been performed on the 
ground-state properties of BN.
All calculations founded on density functional theory (DFT) and on the local
densisy approximation (LDA) for the exchange-correlation potential agree in 
predicting that the c-BN structure has a lower energy than h-BN by about 
0.06eV per atom \cite{furthmueller,albe,wentzcovitch1,wentzcovitch2}.
This result, which is confirmed by our calculation, disagrees with an older 
calculation based on the orthogonal linear combination of atomic orbitals 
(OLCAO), where h-BN was found to be more stable than c-BN by 0.35 eV per atom
\cite{xu}. This is probably explained by a lack of convergence in the older
 calculations.
Also the band structure properties of BN have been the subject of 
%
%Both in our previous DFT-LDA calculation\cite{cappefiore} and 
%in the present paper we found c-BN to be more stable than 
%the h-BN at ambient conditions.
%
%Since the above mentioned cosiderable interest in boron nitride,
extensive thoretical work, performed within different schemes and 
approximations \cite{kleinmann,doni,blase1,rubio,blase2,furthmueller,albe,wentzcovitch1,wentzcovitch2,xu,cappefiore,zunger,tsay,dovesi,robertson,park,catellani,vancamp,suhr,xu2,chrystensen,gavrilenko,cappesatta}.
On the experimental point of view many methods have been employed to 
explore the electronic structure. Among them there are methods like 
soft-x-ray emission spectroscopy (SXRES) \cite{carson,jia}
and photoelectron spectroscopy(PES)\cite{oshima},
which are both sensitive to occupied states, as well as near-edge
x-ray absorption fine structure spectroscopy(NEXAFS)\cite{davies,jimenez} 
and energy loss near edge spectroscopy, which can be used to probe 
the empty bands(ELNES) \cite{mckenzie}.
In addition, electron energy loss spectroscopy performed in the plasmon 
region has proved to provide valuable 
information \cite{mckenzie,buechner,tarrio}.
\par
The experimental optical functions of BN compounds are not 
as well known as those of the other group III nitrides.
This is due in large part to the lack of high quality, single crystal 
samples. Consequently, most optical studies on BN have been performed on 
polycrystalline samples, with various amount of impurities.
As one could expect, the results of optical studies performed on 
these materials vary greatly \cite{miyata,osaka}.
However a general consensus from the experimental data is that the 
minimum band gap is direct in the case of h-BN, and indirect in the case 
of c-BN.
For h-BN a direct band gap of $5.2\pm0.2$eV associated to the transition 
H$_{3v}$-H$_{2c}$ has been 
extimated \cite{hoffman}, while a value of $6.4\pm0.5$eV
for the indirect minimum band gap in c-BN  has been determined \cite{chrenko},
 and associated to the $\Gamma_{15v}$-X$_{1c}$ transition 
\cite{cappefiore,suhr}.
\par
In the present paper we present {\it ab-initio} linear optical 
functions of c- and h-BN studied within DFT-LDA theory. 
No external parameter has been used to fit the experimental curves.  
Due to the importance of the two BN phases, and to the fact that 
in several experimental configurations they appear  
together in the same sample, a parallel study of both phases has been 
accomplished throughout the present paper. 
Complex dielectric functions, refractive index, reflectance spectra, 
optical conductivity, and energy loss functons for c-BN and h-BN have 
been calculated. We report our results in the energy range of interest to
compare them with the existing theoretical and experimental literature.
Calculated DFT-LDA energy-loss functions agree with experiments and 
with previous theoretical results, both for h-BN and c-BN. 
On the other hand, discrepancies arise between theoretical results and 
experiments in the imaginary part of the dielectric function for c-BN. 
Possible explanations of this issue are proposed and evaluated in detail. 
The imaginary part of the dielectric function in the h-BN case, instead, shows 
a reasonable agreement with both other theoretical results and experiments.
\par
This paper is organized as follows. 
In Section II we give the computational details of the calculation of the 
ground state properties for both phases. In Section III we describe  the 
theoretical scheme used to evaluate the optical properties,
and the part related to the calculation of excitation properties. 
In Sections IV and V we present results for the  optical properties  of c-BN 
and h-BN respectively. 
Finally, in Section VI conclusions are drawn, and the  perspectives and issues
opened by our work have been presented.
\section{Computational details and Ground State Calculations}
Density functional calculations have been carried out within the local 
density approximation (LDA) for the exchange and correlation functional
\cite{dreizelgross}, using the Perdew and Zunger parametrization 
\cite{perdewzunger} of the Ceperley and Alder results \cite{ceperlyalder}.
Kohn--Sham orbitals are expanded in a plane--wave
basis set, with an energy cutoff of $55~Ry$. 

Care has been used in constructing the ionic pseudopotentials, in order to 
avoid the occurrence of ghost states and to assure an optimal 
transferability \cite{gonze,fuchs}.
%({\bf Gabriella conferma})
Angular components up to $l=2$ have been  included. 
Separable, norm--conserving soft pseudopotentials have been generated within 
the scheme of Martins and Troullier \cite{troulliermartins}, with the 
following core radii (bohr): 1.59 (B, $3d$), 1.49 (N, $2p$). 
For Boron, non linear core corrections (NLCC) to be used in the solid 
calculation, have been taken into account in generating pseudopotentials and 
pseudowavefunctions \cite{cohennlcc}.
%xxxxxxxxxxxxx
%Discuss the convergency of the band gap energy within DFT-LDA 
%in comparison with other findings({\bf Guido conferma)}).
%xxxxxxxxxxxxxx
\par
With the selected plane-waves cutoff, both the total energy 
and Kohn-Sham eigenvalues are converged to better than 0.1 eV  
for both c-BN and h-BN.
For the cubic phase, the use of 10 special Chadi and Choen $k$ points for 
charge integration in the first Brillouin Zone (BZ) \cite{chadicohen}, is 
found to be sufficient to achieve a good accuracy for the computed total
 energy; for instance, total energy changes by less than 10 meV, and the 
fundamental band gap by less than $0.1~eV$, when passing from 6 to 
10 special points for BZ sampling.
For the layered haxagonal case the use of 12 $k$ points are sufficient 
for a full convergency of the total energy and eigenvalues. 
\par
Hexagonal BN, 
turns out to be less stable than the cubic BN  by about $0.050~ eV/atom$. 
Preliminary calculations in which "hard" norm-conserving pseudopotentials of 
the Bachelet, Hamann, Schl\"uter  (BHS) form were used with energy cutoff of 
150 Rydberg\cite{cappefiore} 
gave an energy difference of $0.078~eV/atom$. 
Our results hence  confirm the previous finding that c-BN is 
more stable than h-BN. 
No contributions to the total energy coming from
zero point vibration of the lattices have been cosidered 
\cite{kern,furthmueller}. 
\par
In Table(I) the structural parameters for the two phases are reported, in 
comparison with previous theoretical calculations and experiments.
For c-BN our equilibrium lattice parameter underestimates the experimental one
by few percent as usual for DFT-LDA, in 
accordance with previous theoretical results. The bulk modulus 
falls very near to the experimental energy range \cite{furthmueller,xu}.
\par
An underestimation of the same order is found for the two lattice parameters 
$a$ and $c\over a$ of h-BN, as reported by  other authors too 
\cite{furthmueller,xu}.

%\end{document}
%***end*
%
No large differences in equilibrium structural 
parameters arise when the harder
BHS pseudopotentials and the larger cutoff are used \cite{cappefiore}.

\section{Optical Properties and Electronic Excitations}
\par
The experimental scenario about the electronic properties of
Boron Nitride is characterized by many measurements. 
For h-BN, however, several old optical and energy loss 
data have been very conflicting each other mostly because of the use of poor 
samples. Band structures calculated in the past were not accurate enough 
to be of real help \cite{xu}.
Recently, careful measurements of inelastic-electron-scattering spectra 
on well characterized h-BN became available \cite{tarrio}, and accurate 
measurements of the linear optical properties of c-BN have been 
reported \cite{miyata}. 
Ellipsometry measurements have also been done to 
characterize noncrystalline thin films of BN \cite{tarrio,aita},
and optical reflectance spectra of c-BN have been recently 
published \cite{osaka}.
\par
We determine the optical properties of h-BN and c-BN by calculating the  
momentum matrix elements (MME) associated with dipole transitions 
at  a large number of k-points in the BZ. 
The frequency-dependent imaginary part of the dielectric 
function is given by\cite{pulci,adolph}
$$
\epsilon_{\alpha \alpha}^{''} (\omega)\,  = \, {(8\pi)^{2} e^2 \over \omega^2 m^2 V}
\sum_{v,c} 
$$
$$
\sum_{\vec k} |<v,\vec k|\,{ p_{\alpha \alpha}}\,| c,\vec k>|^2
\delta (E_c(\vec k)- E_v(\vec k) -\hbar \omega)
$$
\begin{equation}
\label{eq:1}
\end{equation}
where v and c label the valence and conduction states associated with the
energies $E_{v}(\vec k)$, $E_{c}(\vec k)$ , V is the crystal volume, and
$< >$ is the matrix element of the momentum operator.
Nonlocality effects in the ionic pseudopotentials are neglected  
in the evaluation of $\epsilon_2$ \cite{pulci}.
Eigenvalues and eigenfunctions appearing in Eq.(1),
are those determined within the DFT-LDA scheme.
As it appears from Eq.(1), local field effects are also neglected in the
dielectric function calculation \cite{pulci}.
Reference formulas for the macroscopic optical 
functions, namely \cite{baldereschiquattropani}
\begin{equation}
N(\omega)= n(\omega)+ ik(\omega)
\label{eq:2a}
\end{equation}
for the complex refraction index, will be frequently used in the following.
Its relation with the complex macroscopic dielectric function reads:
\begin{equation}
N^2(\omega)= \epsilon(\omega)
\label{eq:2b}
\end{equation}
with
\begin{eqnarray}
\epsilon_1(\omega)=n^2(\omega)-k^2(\omega)
\label{eq:2c}
\end{eqnarray}
and
\begin{eqnarray}
\epsilon_2(\omega)=2n(\omega)k(\omega)\,\, .
\label{eq:2d}
\end{eqnarray}
The absorption coefficient reads (c is the speed of light) 
\begin{eqnarray}
\eta(\omega)= {2\omega k(\omega)/ c}
\label{eq:2e}
\end{eqnarray}
which can be directly connected with
the imaginary part of the dielectric function
\begin{eqnarray}
\eta(\omega) = {\omega \epsilon_2(\omega) / n(\omega) c}
\label{eq:2f}
\end{eqnarray}
 and to the optical conductivity $\sigma$
\begin{eqnarray}
 \sigma(\omega) = {\eta(\omega) n(\omega) c / 4 \pi} 
\label{eq:2g}
\end{eqnarray}
In the above formulas once $\epsilon_2$ (or $k(\omega)$)
is calculated or measured, the correspondent real part 
$\epsilon_1$ (or $n(\omega)$) can be determined via the Kramers-Kroenig 
transformation \cite{baldereschiquattropani}.
\par
\begin{figure}
 \epsfxsize=12.0cm
% \epsfbox{sigmac2.eps}
 \epsfbox{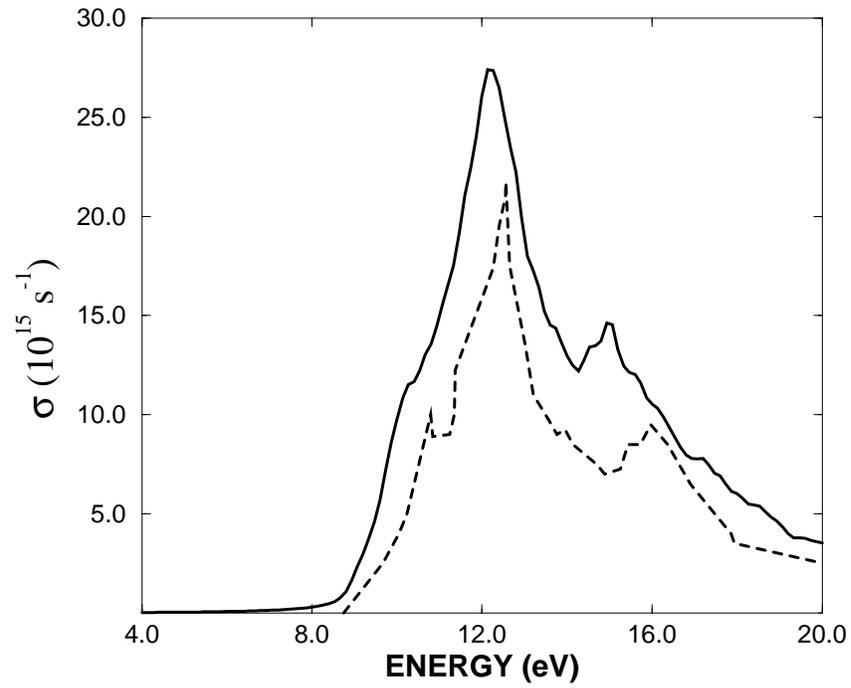}
\vspace*{2.0cm}
 \caption{Calculated optical conductivity $\sigma$ of c-BN (solid line) 
vs. the theoretical result of Ref. \ \protect\cite{xu} (dashed line).}
 \label{sigma}
  \end{figure}
\begin{figure}
 \epsfxsize=12.0cm
 \epsfbox{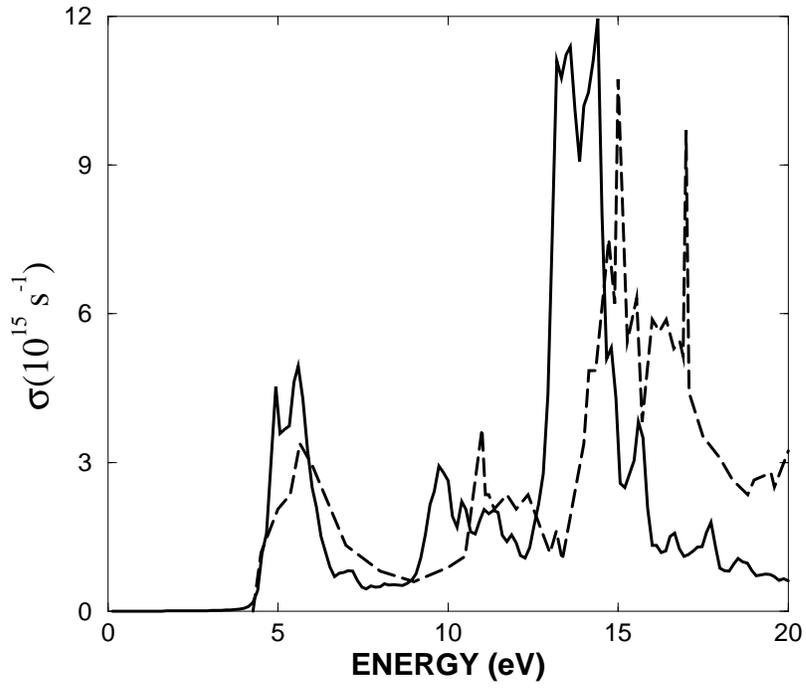}
\vspace*{2.0cm}
 \caption{ Optical conductivity function of h-BN (solid line) vs. 
previous theoretical result (dashed line) after Ref. \protect\cite{xu}.}
% \caption{Calculated optical conductivity function of hexagonal BN}
 \label{sigmahex}
  \end{figure}

To give an overview of the optical properties of the two crystals, and
in particular of the differences among 
the hexagonal and the cubic phase,
 the calculated interband optical conductivity $\sigma$ is reported in 
in Fig. \ref{sigma} and in Fig. \ref{sigmahex}.
In the case of h-BN, an average over the three crystallographics directions is 
taken.
The  Joint Density of States (JDOS) for the same samples are also shown. 
%JDOS  is obtainable from the 
%MME(see Eq(1)) posing equal to one the momentum matrix elements in 
%the calculation.
%The two quantities look different infact, 
Both the absorption threshold and the 
oscillator strengths in h-BN  are much smaller than in c-BN.
In the two cases, our data fairly reproduce the ortogonalized LCAO ones 
by Xu and Ching \cite{xu}.
In both crystals there is 
a very little resemblance between the optical conductivity and the 
JDOS curves, especially in the high frequency region, indicating a strong
$\vec k$ dependence of the MME.
%SONOQUI
Both phases  show a fundamental 
indirect band gap, but in h-BN the indirect gap is smaller than in c-BN
\cite{rubio,cappefiore,suhr}. 
This leads to the lower absorption threshold in h-BN. 
Many structures can be directly related to measured ones, like the 
peak around 6 eV for h-BN which can be assigned to transition between 
bands in the in-plane directions. 
This correspondence reflects the strong anisotropy in the 
optical absorption of this material, which  shows up even when macroscopic
quantities are taken into account (Eq.(2)-Eq.(8)).
\par 
As is well known DFT-LDA band structures for semiconductors and 
insulators cannot reproduce the real (experimental) ones.
DFT-LDA eigenvalues, when interpreted as quasiparticle (QP) energies show the
 so called {\it band gap problem} \cite{hl}.
The QP energies correspond only qualitatively to the
DFT-LDA ones, mainly because the band gaps between conduction and valence 
bands show a systematic underestimation with respect to the 
experiments \cite{bechstedt,bechstedtdelsole,gygi}.
This problem can be solved, for semiconductors and 
insulators, within a self-energy scheme, called GW \cite{hedin}.
Within  this method, the self-energy operator
reads $\Sigma=GW$, where G is the one-electron Green function and 
W the screened Coulomb interaction in the system, 
fully taking into account the screening properties of the 
material \cite{hl,hedin,godby}.
The corresponding Dyson equation can be solved within first-order 
perturbation theory with respect to ($\Sigma-V_{\rm xc}$), 
where $V_{\rm xc}$ is the DFT-LDA exchange and correlation potential.
Assuming the DFT-LDA eigenvalues and eigenfuntions 
as the zeroth order eigenvalues and eigenfuntions of the system respectively,
the QP corrections for the Bloch state $\psi_{n\vec k}$ state 
read \cite{hl}:
\begin{equation}
E_{n\vec k} - E^{(0)}_{n\vec k} = 
{\Sigma^{\rm coh}_{n \vec k}(E^{(0)}_{n\vec k})+
\Sigma^{\rm sex}_{n \vec k}(E^{(0)}_{n\vec k})- 
\,\langle \psi_{n \vec k} | V_{\rm xc} | \psi_{n \vec k} \rangle 
\over{1+d\Sigma_{n\vec k}(E)/d E|_{E^{(0)}_{n\vec k}} }}
%\label{eq:3}
\end{equation}
where $\Sigma^{\rm coh}_{n \vec k}$, $\Sigma^{\rm sex}_{n \vec k}$ and
$V^{LDA}_{xc,n\vec k}$ are the expectation values of the
Coulomb Hole (COH), Screened Exchange (SEX) 
contributions to the self energy, and of the DFT-LDA exchange-correlation
(XC) potential, respectively.\cite{hl}
{\it Ab-initio} GW schemes are based on the full calculation
of the screening function of the system 
(i.e. fully including local field and dynamical effects)
starting from  DFT-LDA eigenvalues and eigenfunctions. 
These kind of calculations for semiconductors and insulators have often led to
a very good agreement between theory and experiment \cite{hl,godby}.
\par
Self-energy corrections to the DFT-LDA band structure for semiconductors and 
insulators can also be performed, with very good results, also using a model 
dielectric function to mimic the real screening of the 
system \cite{bechstedt,bechstedtdelsole,gygi}.
One can start for a perturbative formula of the form:\cite{bechstedt92}
\begin{equation}
E_{n\vec k}- E^{(0)}_{n\vec k} =  
{\Sigma^{\rm coh}_{n \vec k}+\Sigma^{\rm sex}_{n \vec k}+
\Sigma^{\rm dyn}_{n \vec k}(E^{(0)}_{n\vec k})-
\,\langle \psi_{n \vec k} | V_{\rm xc} | \psi_{n \vec k} \rangle 
\over {1+d\Sigma_{n\vec k}(E)/dE|_{E^{(0)}_{n\vec k}} }}
%\label{eq:4}
\end{equation}
where $\Sigma^{\rm dyn}_{n \vec k}(E^{(0)}_{n\vec k})$ is 
the zeroth order term of the energy expansion of the dynamical part 
of $\Sigma$, and only the static parts of the SEX and COH self-energy are 
present \cite{hl,bechstedt92}.
Corrections to the DFT-LDA spectra calculated from Eq.(10) 
with the use of model dielectric functions
enabled both us and other authors 
to obtain very positive results, when compared with 
experiments and {\it ab-initio} GW ones, in several  
semiconductors and insulators \cite{cappelist,hl88}.
Moreover, the use of model screening functions enables one 
to perform GW calculation in a more efficient way, 
reducing the computational effort  of at least one order of magnitude in CPU 
time \cite{bechstedt92,hl88}.
Details concerning the evaluation of the different coefficient
appearing in Eq.(10) and how do they compare with 
the {\it ab-initio} calculated ones are  given 
elsewhere \cite{cappelist,cappefrancesi}.
\par 
Either {\it ab-initio} or simplified  GW schemes have been 
employed to calculate the QP energies of h-BN and 
c-BN \cite{cappefiore,suhr,cappesatta,blase}.
All theoretical results improve the agreement with experimental data. 
Tipically, LDA transition energies suffer of an undersetimation of
about  3eV for the cubic phase, and 2eV for the hexagonal one 
\cite{cappefiore}. 
In Tab.(II) and Tab.(III)  for c-BN and h-BN respectively
we report the results of previous 
calculations and experiments, namely  the fundamental gaps and the main 
transition energies, in comparison with present results. 
In c-BN the present transition energies have been calculated at the 
corresponding theoretical lattice costant while in ref. \cite{suhr}, (as
 explicitely stated by the authors) the experimental lattice parameter
has been used. Moreover, the use of an efficient method in a large gap 
system induces a slight overestimation of the above 
energies \cite{cappefiore,cappefrancesi}.
As it is clear from the data, minor differences arise between the 
{\it ab-initio} GW results and the efficient GW ones.
%***********exzcitatin table
%
%

In Tab.(III), the value marked with "a" correspond to a 
transition between a point T$_1$ (along the $\Gamma-K$ direction near K, and 
near in energy to H$_v$) and M$_c$. Similarly the term with "b" 
is the transition energy between K$_{3v}$-K$_{3c}$, which is near in energy 
to the transition H$_{3v}$-H$_{2c}$ \cite{suhr}.

\section{Cubic BN}
\begin{figure}
 \epsfxsize=12.0cm
% \epsfbox{twocomp.ps}
 \epsfbox{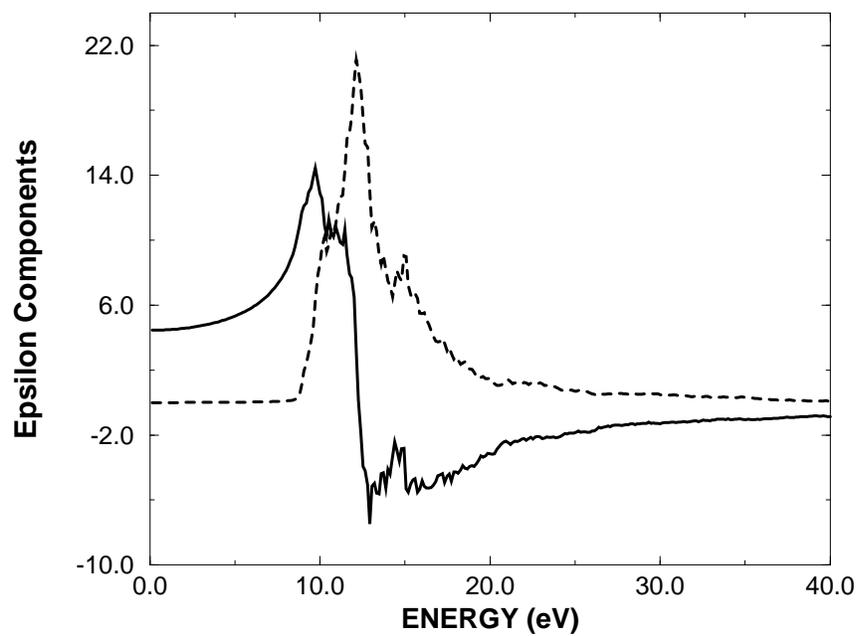}
 \vspace*{2.0cm}
 \caption{ Computed real and imaginary parts of the  dielectric function for
 c-BN.
   $\epsilon_1$ is the solid line, $\epsilon_2$ the dashed one.}
 \label{twocomp}
 \end{figure}

In Fig. \ref{twocomp} we report the calculated dielectric function for 
c-BN obtained within DFT-LDA. The static dielectric constant is found to be 
of 5.45, larger than the experimental one (4.45) \cite{madelung}. 
This fact must be ascribed to absence of local fields effects in the 
calculation of $\epsilon_2$ \cite{gavrilenko,adolph}.
Experimental curves obtained from  
recent optical reflectance measurements  in the energy range $5\div25$ eV 
with syncrotron radiation by Osaka e coworkers \cite{osaka} are also reported.
Two different samples have been used by them, namely
a sintered c-BN plate and c-BN thin film.
The c-BN sintered plate produced at high pressure and temperature, 
offered by Sumimoto Electric Industry (Japan), had dimension 
of 5x5x0.5mm$^3$ and was characterized 
as cubic by using X-ray diffraction and Raman Scattering \cite{osaka}. 
The film, on the other hand, was grown using chemical vapor deposition 
(CVD) on a silicon substrate, with negative self-bias. 
c-BN phase, syntesized from $B_2H_6$ (Ar diluition)
and $N_2$ gas mixtures, had a the grain size is 100-200 \AA, 
controlled by Transmission Electron Microscopy (TEM) images. 
\par
\begin{figure}
 \epsfxsize=12.0cm
% \epsfbox{reflectance.ps}
 \epsfbox{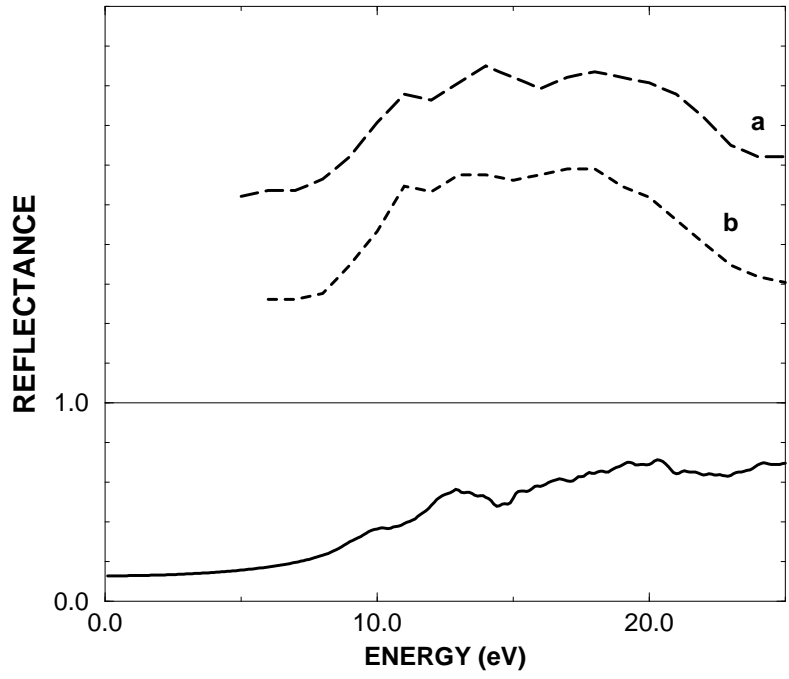}
\vspace*{2.0cm}
 \caption{Calculated reflectivity (solid line) of c-BN vs. experimental 
results in arbitrary units (dashed lines), taken from  Ref. \protect\cite{osaka}.}
 \label{reflec}
  \end{figure}
\begin{figure}
 \epsfxsize=12.0cm
 \epsfbox{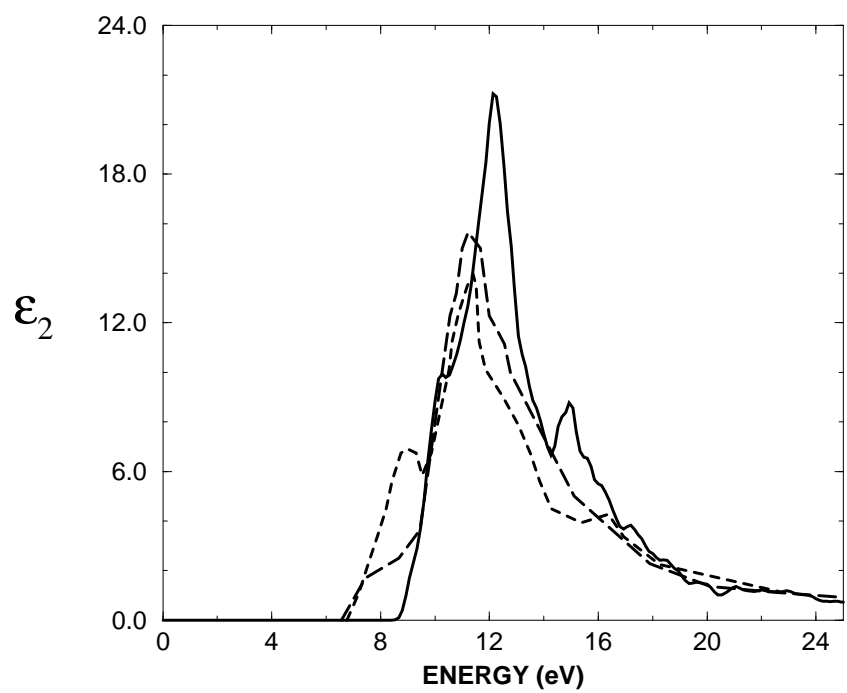}
\vspace*{2.0cm}
 \caption{Calculated imaginary part of dielectric function of c-BN 
(solid line) vs. the experimental results of Ref. \protect\cite{osaka}
 ( dashed lines).}
 \label{cfrepsilon}
  \end{figure}
\begin{figure}
 \epsfxsize=12.0cm
 \epsfbox{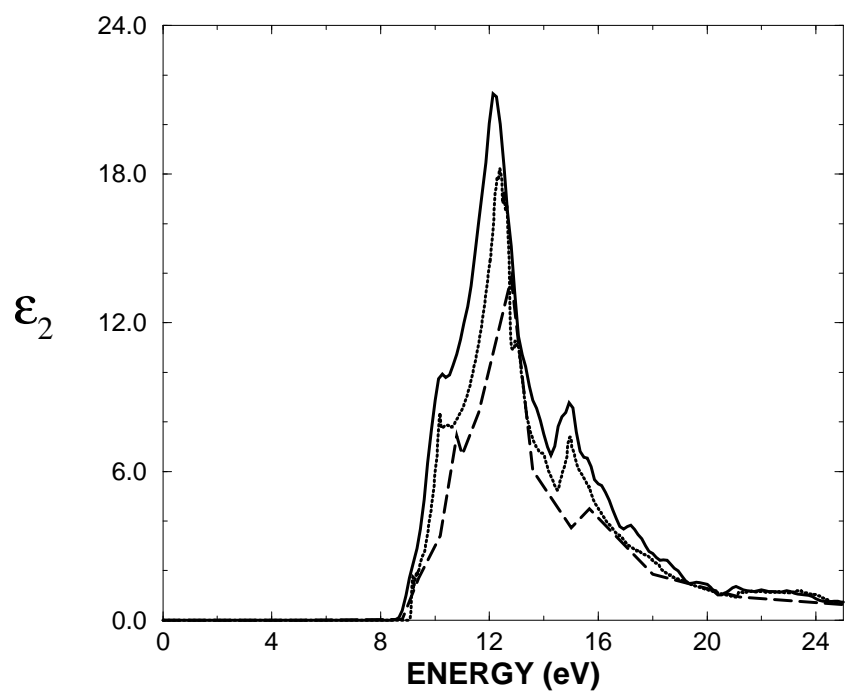}
\vspace*{2.0cm}
 \caption{Calculated imaginary part of dielectric function of c-BN 
(solid line) vs. LMTO result after Ref.\protect\cite{monachesi} (dotted line) 
and LCAO result (dashed line) of Ref.\protect\cite{xu}.}
 \label{monach}
  \end{figure}

In Fig. \ref{reflec} the calculated reflectivity of c-BN is displayed together
 with the experimental data taken from ref. \cite{osaka}. 
Both the spectra show peaks at 11.7 and 14.0eV and a broad 
structure around 18eV.
The authors of ref. \cite{osaka} claim that the first two peaks correspond 
to the E1 and E2 peaks of the zinc-blende-type semiconductors \cite{madelung}.
Our theoretical curve shows  two structures centered at 
$12.8~eV$ and $14.4~eV$  but no evidence of higher energy peaks 
results.
To enter in further detail with other theoretical data and experiments, 
we report in Fig. \ref{cfrepsilon} the imaginary part of $\epsilon_2$, in 
comparison with two experimental curves by Osaka et al. \cite{osaka}
One of these curves has been obtained from a crystalline sample, 
and the other from a film grown by CVD technique. In Fig. \ref{monach} 
we also compare our curve with that calculated by Xu and Ching \cite{xu}, and 
with a LMTO one calculated at our theoretical lattice parameter 
\cite{monachesi}.
It is significant to discuss these figures in detail. 
The experimental data show major structures A and B at 
9.05 and 11.7 eV, and shoulders at 13.2eV, 16.7eV (C and D) for the
crystalline sample.  
In the $\epsilon_2$ in the c-BN film, the peak A and the 
shoulders C and D are not present.
The spectrum of the c-BN thin film deposited by plasma CVD is 
similar to that of the c-BN synthesized under high pressure and temperature.
The authors claim that the differences within their spectra
is a demonstration of the fact that the microcrystalline c-BN does not 
grow epitaxially on silicon substrate, and presents a rough surface. 
They do not mention the value of the absorption onset, but a value of $6.8~eV$ 
can be deduced from their figure of $\epsilon_2$. 
To consider the correspondence with the electronic transitions, peak A has
been associated to the transition $\Gamma_{15v}-\Gamma_{15c}$,
the transition X$_{5v}$-X$_{3c}$ and X$_{5v}$-X$_{1c}$, peak C to the 
transition L$_{3v}$-L$_{3c}$, and peak D at 16.7 eV  
to L$_{1v}$-L$_{1c}$ \cite{osaka}.
The present DFT-LDA reflectance spectrum reproduced qualitatively well the 
experimental curves, but the first peak is found at 12.8eV and the following  
at $15.2~eV$. 
Fig. \ref{monach} also demonstrates that our results do agree with  
 other theoretical calculations (moreover the theoretical curves reported show 
no remarkable difference in the range $25-40~eV$ which has not been showed in 
this figure). 
However, the absorption  onset obtained within DFT-LDA is at $8.9~eV$, i.e.
much {\it higher} that the experimental one. 
This issue turns out to be rather singular
in sight to the the fact that usually , due to the above mentioned 
{\it band gap problem},
the DFT-LDA calculated absorption onset should fall at {\it lower} energies
%show an onset at lower energies 
than the experimental ones \cite{chrystensen,gavrilenko,xu,hl}. 
In our case the LDA theoretical curve is blueshifted by more than 2eV
with respect to the experiment. A similar overstimate is found also  by 
other theoretical calculations \cite{xu,gavrilenko,chrystensen}.
The interpretation of Christensen and Gorczyca \cite{chrystensen}
for $\epsilon_2$ is different (away from treshold) from
 that proposed by Osaka e 
coworkers \cite{osaka}, Xu and Ching \cite{xu} and from our interpretation.
 Infact the authors of ref.\cite{chrystensen} associate the A peak
to the  transition 3 to 5 at the $1/2\Gamma U$ point, 
the major peak B to the  4-5 transition along $\Sigma$ and L point, the C 
peak to the 4-6 transitions at L, and the D peak to transition 4-6 at $\Delta$
 and at $\Sigma'$. The onset is ascribed to direct transition at 
$\Gamma$. Xu and Ching found A, B and D structures respectively at 
10.7eV, 12.6eV and 15.6 eV, but no evidence for the C shoulder in their curve 
results \cite{xu}. Also their data show a rigid blueshift with respect to the 
experimental curve by about 1.7eV.
Our results rensemble those of Xu and Ching.
Please notice that their name assignements for the spectrum does not match 
the present ones. 
Our onset is found at 9.05eV, while the structures A, B, C and D are 
found respectively at $9.7~eV$, $12.5~eV$, $14.9~eV$, with small evidence of 
peak D.
Our assignement to band transitions within DFT-LDA agrees with that 
proposed in Ref.\cite{osaka}, which has also been confirmed by 
Tsay and coworkers \cite{tsay}.
\par
As far as we know the only other existing experimental data for the 
minimal direct gap of c-BN are those by Philipps and Taft \cite{philiz},
which yield a value of 14.5 eV. 
They studied diamond in the vacuum ultraviolet, and they reported data about 
a c-BN sample displaying structures around 9 and 10 eV, and a peak near 
14.5 eV \cite{philiz}.
The BN data were reported as marginal ones by the authors, to support the 
evidence for a larger direct band gap in c-BN than in diamond. Their value of 
14.5eV has been considered  as the experimental direct band gap of this 
material by other authors in the literature \cite{suhr,madelung}.
On the other hand, these data could be interpreted in good agreement with our 
DFT-LDA spectra. Infact, one could consider the first structure as the display
of the onset (at 9eV in our DFT-LDA calculation) 
and the major peak as the maximum (at 12.5 eV in our DFT-LDA calculation) 
(see Fig.s(2)-(4)).
\par
\begin{figure}
 \epsfxsize=12.0cm
 \epsfbox{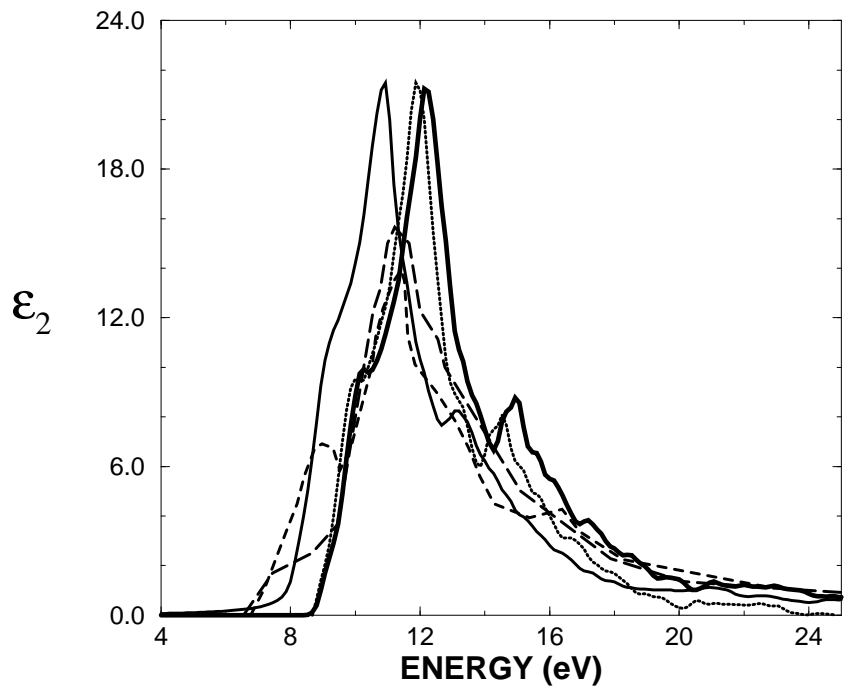}
\vspace*{2.0cm}
 \caption{Calculated imaginary part of dielectric function of c-BN at  
experimental (dotted line) , theoretical (bold solid line) and theoretical 
plus $5\%$ (solid line) lattice constant.
 Experimental results (dashed lines) after Ref. \protect\cite{osaka} are
 plotted for reference.}
 \label{testeps2}
  \end{figure}

Let us now address in detail the main mismatch 
between theoretical results (including ours) and the experiments, concerning 
the onset.
One of the points to be considerd is the problem of 
the correct value of the lattice parameter to be used in optical properties
calculations. This issue has been raised for BN 
by experimental and theoretical works \cite{osaka,gavrilenko}.
In Fig.(\ref{testeps2}) we reported different DFT-LDA curves for 
$\epsilon_2$ calculated at different lattice parameters around the equilibrium
one. It is evident that the onset of the theoretical 
absorption meets the experimental one only when the lattice 
constant is expandend, with respect to the equilibrium one, by $5\%$. 
This value appear to be too large, both with respect to the experimental one, 
and in sight of the fact that:
I) usually the DFT-LDA understimation of the lattice costant is about    
($1\%$) \cite{palummo,fiorentini}, and II) usually the DFT-LDA 
{\it understimates} the absorption onset.

Another possibility is that the sample under experimental study has 
not been carefully characterized, namely concerning different  possibilities 
of disturbances, as thermal ones, or the presence of (oxide) overlayers 
\cite{gavrilenko,osaka}, 
which we will not address here but which cannot be excluded in principle.
In the following,  we consider instead the possibility of h-BN contamination 
within the c-BN sample. 
This is possible due to the fact that in normal conditions 
h-BN domains may be found in c-BN samples \cite{widmayer}.
\par
\begin{figure}
 \epsfxsize=12.0cm
% \epsfbox{n.ps}
 \epsfbox{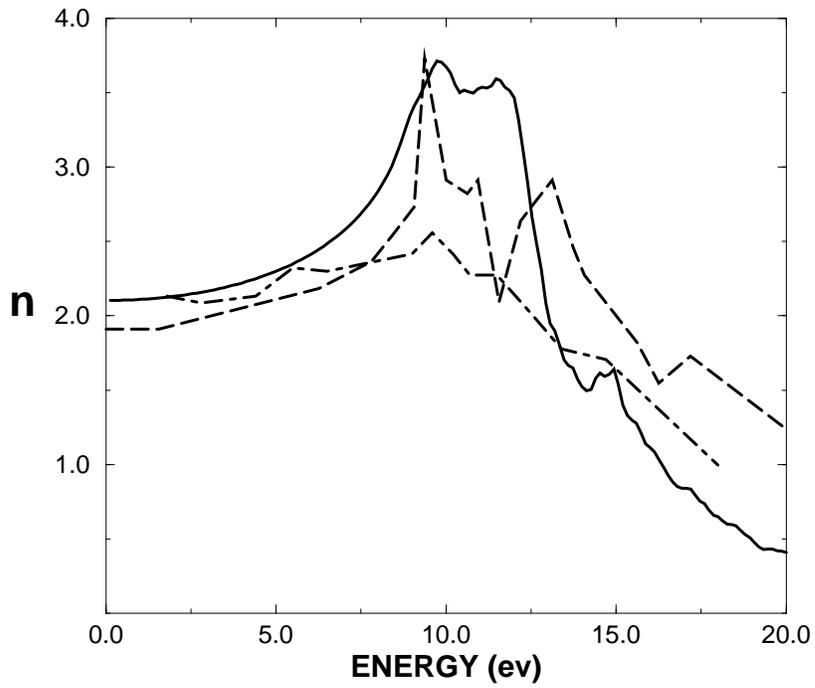}
\vspace*{2.0cm}
 \caption{Calculated real part of refracive index of c-BN (solid line) 
vs. the experimental results (dot-dashed line after Ref.\protect\cite{miyata})
, and previous theoretical result (dashed line after Ref.\protect\cite{xu}).}
 \label{npart}
  \end{figure}
\begin{figure}
 \epsfxsize=12.0cm
 \epsfbox{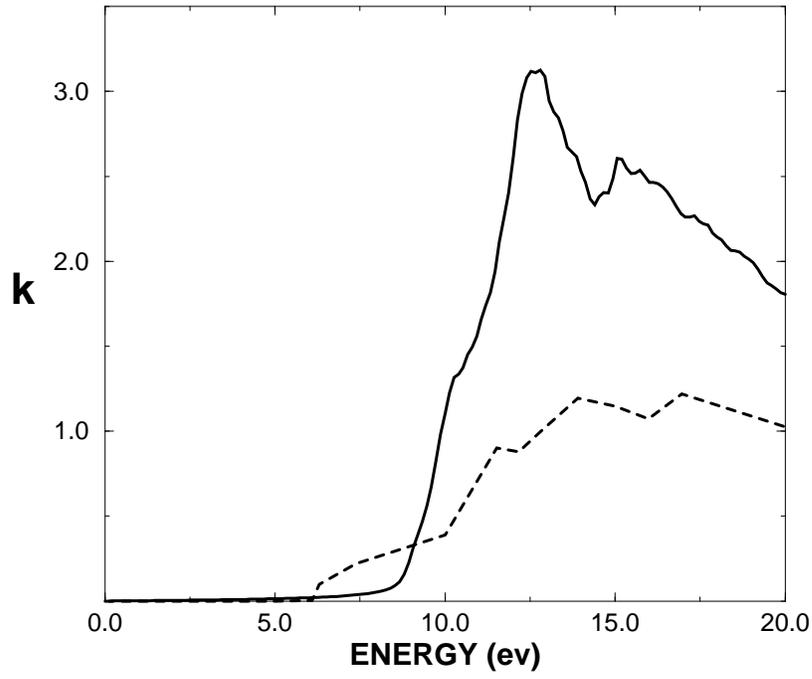}
\vspace*{2.0cm}
 \caption{Calculated imaginary  part of refracive index  of c-BN (solid 
 line) vs. experimental results (dashed line) after Ref.\protect\cite{miyata}.}
 \label{kpart}
  \end{figure}

%DA CORREGGERE
In Fig. \ref{npart} and Fig. \ref{kpart} we also report the real and imaginary
 part of the refractive index, $N(\omega)$ and the reflectance spectrum in 
comparison with the experimental data of Miyata and coworkers \cite{miyata}.
The experimental data have been obtained from reflectance measurements over a 
photon energy range between $2\div23$eV, and from transmittance data in the 
range  $2\div7$eV.
A single crystal of c-BN (5mm$^2$ area x 0.16mm thick) has been used. 
Both the onset of the absorption coefficient, or the imaginary part of the 
refractive index, indicate a gap of $6.1\pm0.5eV$. Our results are only in 
qualitative agreement with  these  measeurements.

We can argue that part of the error comes from the normalization imposed
to the experimental curves. 
Infact, the absolute value of reflectance and transmittance have been obtained
by the authors of ref. \cite{miyata} by imposing that refractive index at 
$\omega=2.10~eV$ equal to 2.117, the value determined by an indipendent 
experiment by Gielisser et al. \cite{gielisser}.
Within the present work instead, no adjustable parameter has been used to fit 
the experimental data.
\par
\begin{figure}
 \epsfxsize=12.0cm
% \epsfbox{elosscub.eps}
 \epsfbox{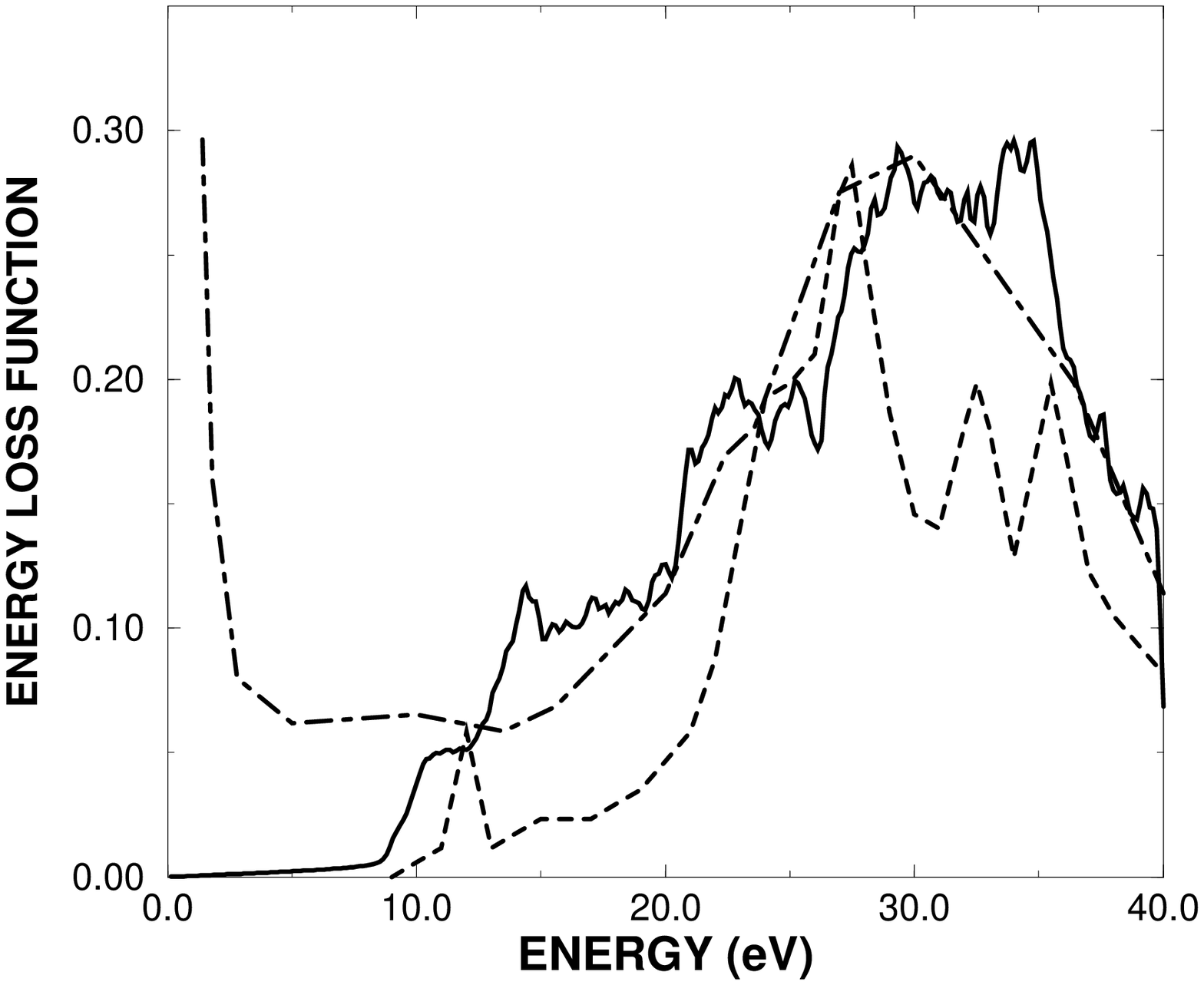}
\vspace*{2.0cm}
 \caption{ Energy-loss function of c-BN (solid line) vs. the experimental
 result (dot-dashed line) after Ref.\ \protect\cite{mckenzie} and previous 
theoretical result (dashed line) of Ref.\ \protect\cite{xu}.}
% \caption{Calculated energy loss function of cubic BN (solid line) vs.
% experimental result (dotted line) of Ref. [Mc kenzie]}
 \label{loss}
  \end{figure}

In Fig. \ref{loss} we report the calculated DFT-LDA energy loss function 
for cubic BN , namely
$Im[-1/\epsilon(q,\omega)]$, in comparison with the experiments.
We compare our results with  Electron Energy Loss (EEL) mesaurements 
of McKenzie et al. \cite{mckenzie}. 
The dominant maximum in the EEL spectrum is due to the plasmon excitation, a 
longitudinal oscillation of the valence electrons as a whole against the 
cores,classically occurring at frequency  $\omega^2_p= N e^2/m $  were N 
represents the density of the valence electrons of the sample. 
Due to the different density of h-BN and c-BN, EEL spectra have been used to 
discriminate between the two phases within the same specimen \cite{mckenzie}.
The measured $\omega_p$ for h-BN is 25.5eV and for the c-BN is 28.5eV. 
An absorption threshold for the cubic phase at 9eV has been deduced by the 
authors of ref. \cite{mckenzie}.
Our DFT-LDA calculation yields  a first peak at 30eV, which is in 
reasonable agreement with the experimental one ($28.5~eV$) and with that 
determined by Xu and Ching (28eV) \cite{xu}.
\section{Layered Hexagonal BN}
\begin{figure}
 \epsfxsize=12.0cm
% \epsfbox{elosshex.eps}
 \epsfbox{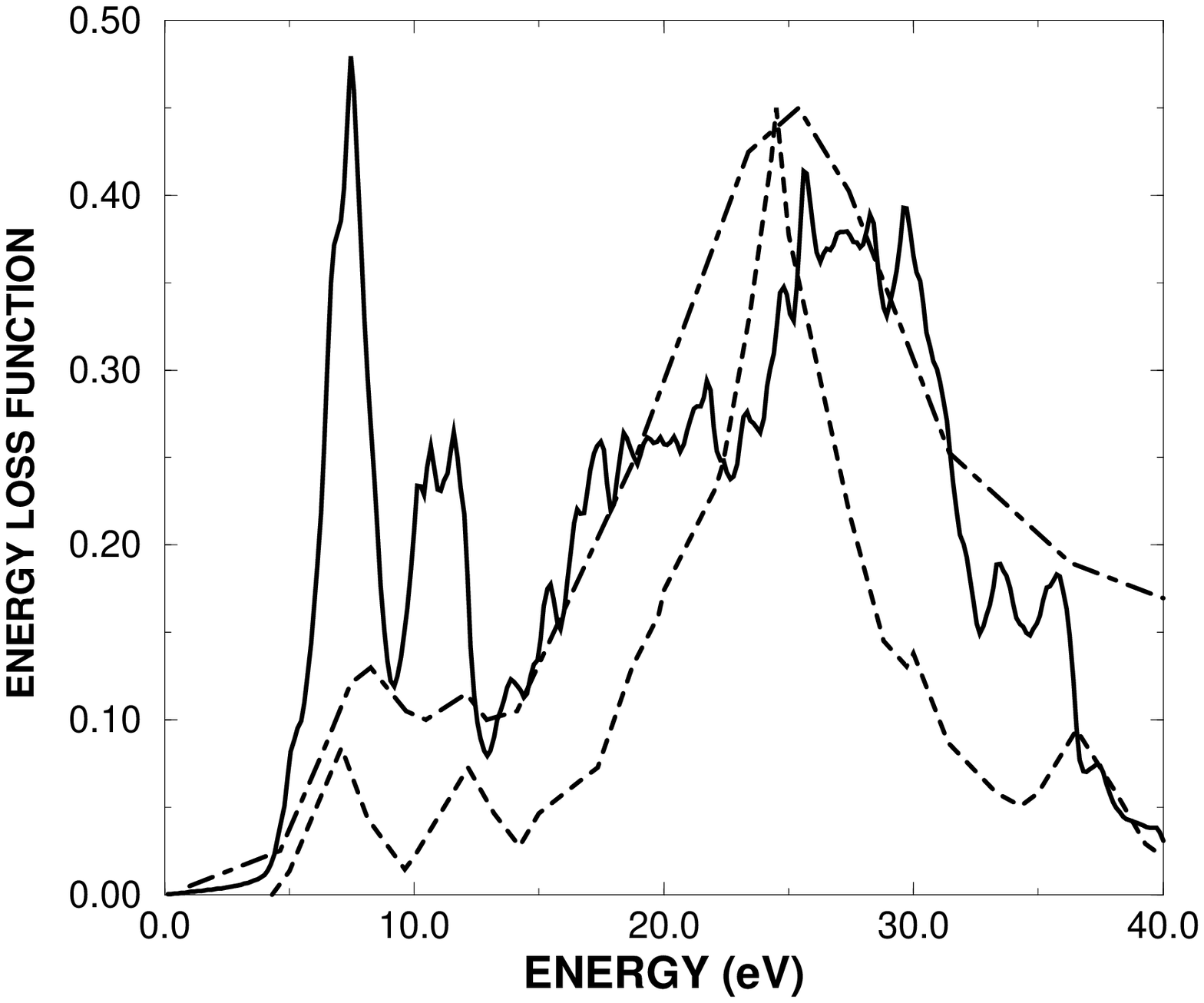}
\vspace*{2.0cm}
 \caption{ Energy-loss function of h-BN (solid line) vs. the 
experimental result (dot-dashed line) of Ref.\ \protect\cite{mckenzie} and 
previous theoretical result (dashed line) after Ref.\protect\cite{xu}.}
 \label{losshex}
  \end{figure}

\begin{figure}
 \epsfxsize=12.0cm
 \epsfbox{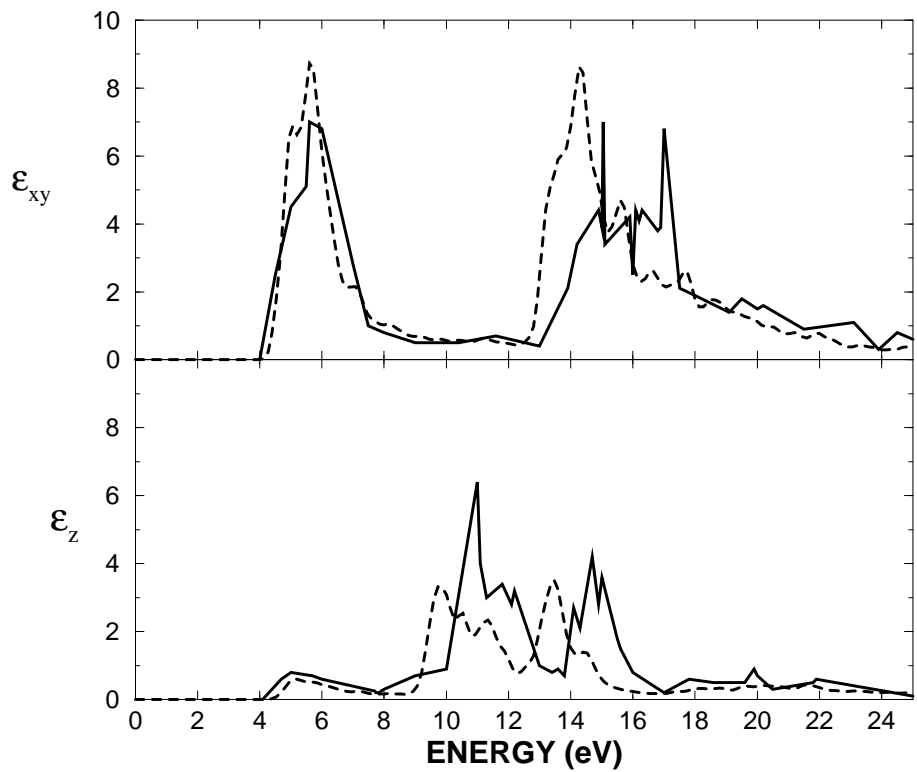}
\vspace*{2.0cm}
\caption{ First panel: imaginary part of the  dielectric function of h-BN 
(solid line) averaged on x and y directions vs. the same quantity after 
Ref. \protect\cite{xu} (dashed line).
Second panel: imaginary part of the  dielectric function of h-BN 
(solid line) on z direction vs. the same quantity after Ref.\protect\cite{xu} 
(dashed line). }
 \label{grafcomphex}
  \end{figure}

\begin{figure}
 \epsfxsize=12.0cm
 \epsfbox{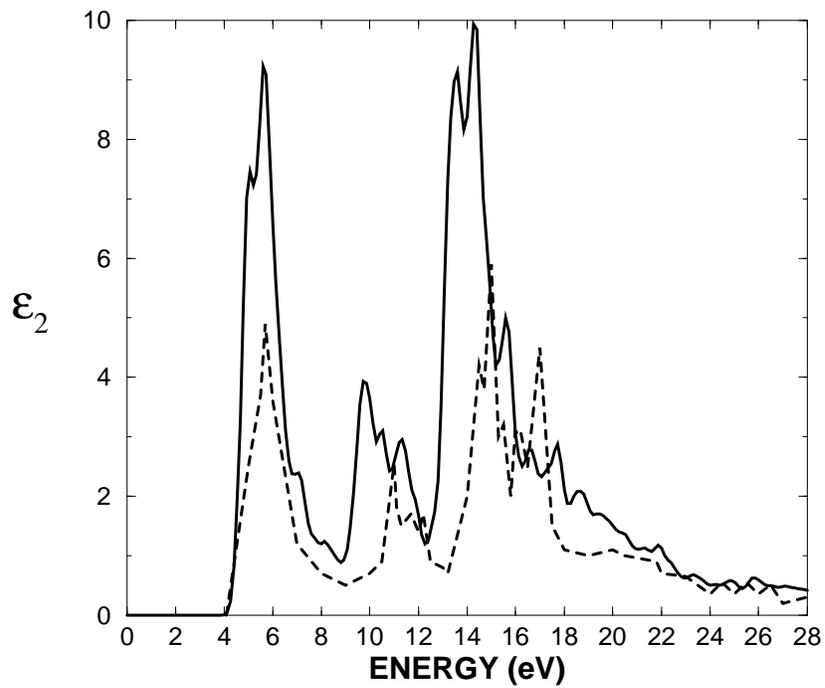}
\vspace*{2.0cm}
\caption{ Imaginary part of dielectric function of h-BN (solid line) vs.
 previous  theoretical result of Ref.\protect\cite{xu} (dashed line) 
averaged over the three crystallographic directions. }
 \label{grafnewhex}
  \end{figure}

\begin{figure}
 \epsfxsize=12.0cm
 \epsfbox{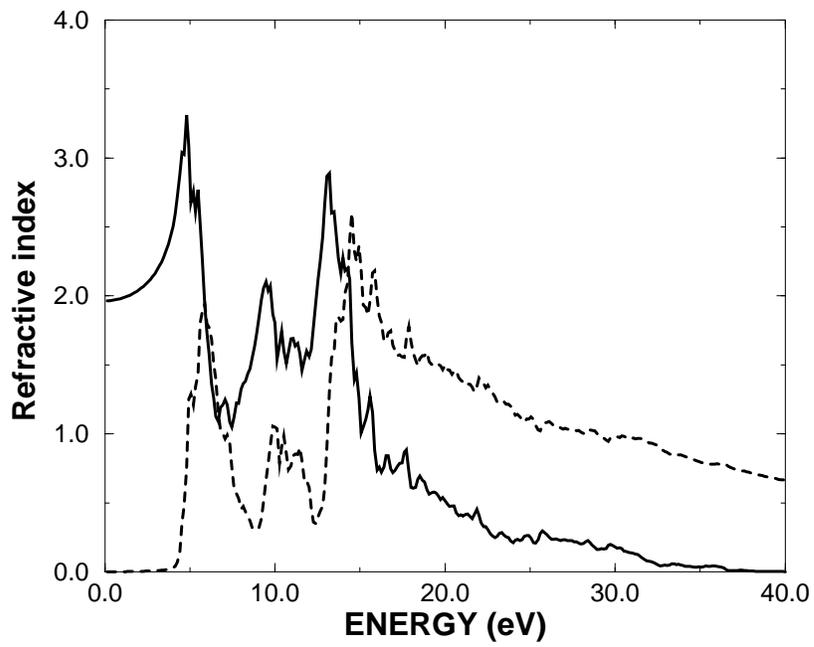}
\vspace*{2.0cm}
 \caption{Calculated refractive index of h-BN: real (solid 
line) and imaginary part (dotted line).}
 \label{indexhex}
  \end{figure}

For this material conflicting experimental and theoretical values of the 
fundamental band gap can be found in the literature.
Tarrio e Schatterly presented EELS spectra 
between $0\div60$ eV, with a plasmon peak at 26.4eV. From their data,
a direct gap of $5.9\pm0.2~eV$ was inferred \cite{tarrio}.
Catellani and coworkers 
using a linear augmented plan wave computational scheme
(FLAPW-LDA), claimed that h-BN has a minimal indirect gap at the 
$H_{3v}-M_{1c}$ transition, corresponding to $3.9~eV$, and a minimal
direct gap at $4.3~eV$, corresponding to the  $H_{3v}-H_{2c}$ transition 
\cite{catellani}.
Hoffman and coworkers, by optical reflectivity measurements in the range 
$0.045\div10~eV$, found a direct gap of $5.2\pm0.2~eV$,  
deducing that it correspond to direct transitions at $H$ \cite{hoffman}.
 Park, Terakura, Hamada, using a  FLAPW (LDA) scheme, found 
a minimal direct gap, and gave
for the $M_v-M_c$ transition energy a value of $4.5~eV$ \cite{park}.
Last but not least, the work of Suhr and coworkers and \cite{suhr}
of Cappellini et al. found, within DFT-GW,
that the material has an indirect minimal gap (see section III).
For h-BN we start the discussion from the EEL function reported 
in Fig.\ref{losshex}. 
In this figure, we compare the calculated DFT-LDA curve, 
averaged over the three crystallographics axis, with the curve 
 calculated by Xu and Ching \cite{xu}, and with the 
experimental one by McKenzie \cite{mckenzie}.
We recognize three major structures.
In the theoretical spectrum of ref.\cite{xu}, a first  peak(A) around 7eV 
can be found, a second peak (B) is at $12eV~$, and a major 
structure (C) appears at $24~eV$.
The latter is the bulk plasmon peak, to be compared with 
the experimental value of 25.5 eV \cite{mckenzie}.
%%%%%%%%%
%{\bf DA RIVEDERE}
%%%%%%%%%%%%
In Fig. \ref{grafcomphex} we report the calculated 
imaginary part of the dielectric function for x-y and z components.
% averaged over the three crystallographics axis. 
Our results compare  well also in this case with the 
LCAO ones by Xu and Ching. In Fig. \ref{grafnewhex} we plot the imaginary part
 of the dielectric function averaged over the three crystallographics axis.
In particular after Fig. \ref{grafcomphex}  and Fig. \ref{grafnewhex}, 
we agree with the conclusions of Xu and Ching that peak B in the EEL function 
of  Fig.\ref{losshex},
comes from the component of $\epsilon_2$ parallel to the c axis of the 
hexagonal crystal, while
peak A comes form the perpendicular components of $\epsilon_2$. 
These statements may be easily confirmed by considering experimental curves
at different scattering momentum transfer (in plane versus c-axis data) 
\cite{xu,tarrio}.
%%%%%%%%%%%%%%%%%%%%%%
The major peak C contains contributions from both  the $\epsilon_2$
components perpendicular and parallel to the c-axis.
In Fig. \ref{indexhex} we report the imaginary and real part of the 
calculated refractive index.
Again, a qualitative difference with respect with the cubic case is observable.
Infact, although the imaginary part of the dielectric function has been
averaged over the three crystallographics directions, 
strong differences arises with respect to the cubic case (see Fig.(8)). 
%\par
% Fig.\ref{reflechex} shows the reflectivity spectrum from the present 
%calculations versus the experimental one by Zunger and coworkers 
%\cite{zunger}.
%The comparison is only possible in the first region of the calculated 
%spectrum, where the experimental data were available. 
%Others experiments by Hoffman, Doll, Eklund,\cite{hoffman} extending  to 
%10eV..........({\bf Guido, fare confronto sulla figura}).
%\par

\begin{figure}
 \epsfxsize=12.0cm
 \epsfbox{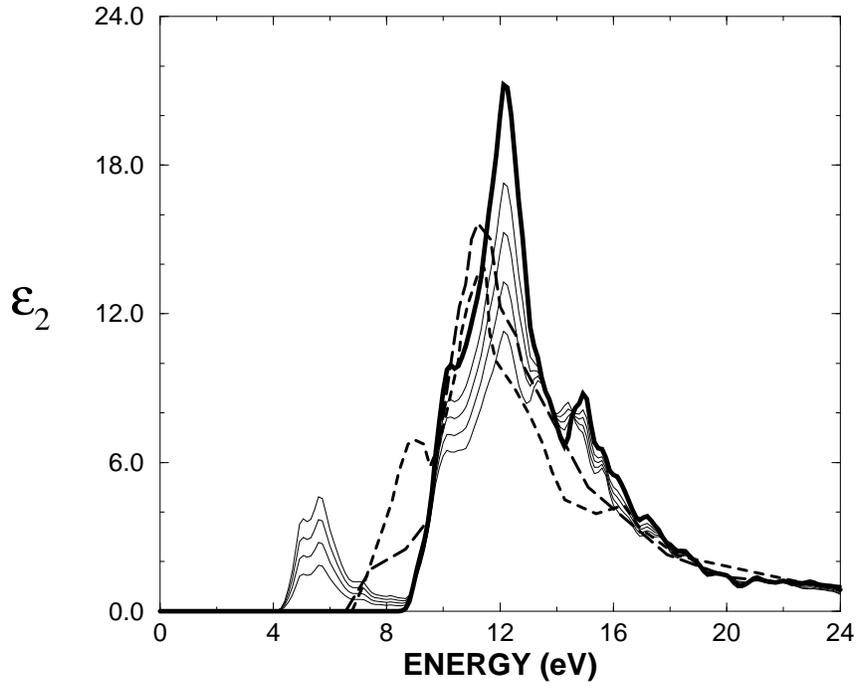}
\vspace*{2.0cm}
 \caption{Imaginary part of dielectric function for merged  c-BN and h-BN 
(solid lines). The experimental  results of Ref.\protect\cite{osaka} are 
also given (dashed lines). Pure cubic $\epsilon_2$ is showed with bold solid 
line.}
 \label{allmed}
  \end{figure}

\begin{figure}
 \epsfxsize=12.0cm
 \epsfbox{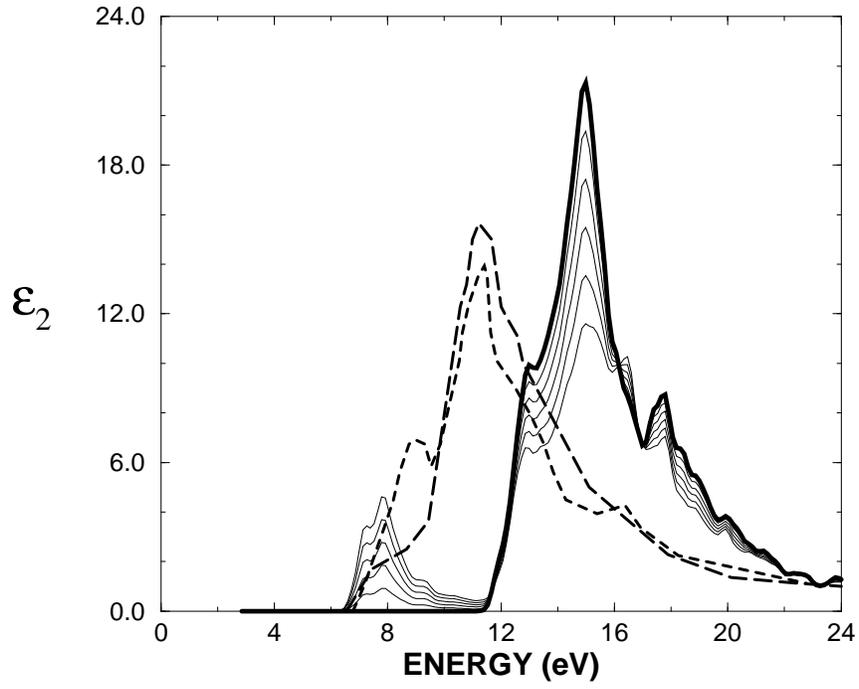}
\vspace*{2.0cm}
 \caption{Imaginary part of dielectric function with GW corrections 
(see text) for merged  c-BN and h-BN 
(solid lines). Experimental  results of Ref.\protect\cite{osaka} are also 
given (dashed lines). Pure cubic $\epsilon_2$ is showed with bold solid line.}
 \label{allmedGW}
\end{figure}

The knowledge of the optical properties of h-BN can help us to understand the
 mismatch between theory and experiments for the $\epsilon_2$ of c-BN.
Infact, as it was recently shown \cite{widmayer}, c-BN films display an 
hexagonal-like top layer, 
and in presence of disturbances the amount of disorder in both types of films 
increases significantly, leading to the transformation of the cubic phase to 
the hexagonal-like material. 
Moreover, it has been demonstrated that due to the large hysteresis of the
hexagonal-cubic phase transitions, h-BN domains may continue to exist in 
c-BN sample \cite{eremets}. 
It can hence be argued that the experimental samples of c-BN are likely to 
contain impurities due to h-BN domains.
Assuming that the sample used in the experimental work by Osaka and 
coworkers \cite{osaka} 
contained h-BN domains, we can work within the effective crystal approximation,
i.e. assuming that the domains are homogeneous and isotropic: in this case 
an imaginary dielectric function corresponding to a linear combination of the 
two pure forms can be expected for the merged system\cite{effective}:
\begin{equation}
\epsilon_2(\omega)\,= x \epsilon_2^{h-BN}(\omega)
+(1-x)\epsilon_2^{c-BN}(\omega)
\end{equation}
with x going from 0 to 0.5.
In Fig. \ref{allmed} we report the results obtained within this scheme always 
at the DFT-LDA level. 
By increasing the concentration of h-phase, a peak at 
$5~eV$ due to h-BN presence does grow, while the peak 
(for which the cubic phase is responsable) around 12eV, due to the cubic phase
 decreases.
Even if the main peak comes in better agreement with the experiments
in intensity, the peak around 5eV remains significantly red 
shifted with respect with the experimental onset.
To address more deeply this issue, we calculate  self-energy corrections 
to the spectra of both phases within the 
GW approximation \cite{cappefiore}. The results are reported in Fig.
\ref{allmedGW}.
The GW corrected spectrum matches the 
experimental treshold better than the DFT-LDA one, but the major peak now 
misses completely the main experimental structure at 11.7eV.
\par
Another possible explanation to the mismatch between theory an experiment 
for the cubic phase  is the contribution  of 
higher-order effects in the spectra, like those due to the electron-hole 
interaction which we have neglected so far.
If we consider, in Fig. \ref{allmedGW}, the case with x=0 (pure c-BN with GW 
self-energy corrections), the GW theoretical onset falls at 
11.6eV (see section 1), more than 5eV higher than the experimental 
one ($\sim 5.8eV$ by Osaka). 
Hence the inclusion of self-energy effects in the DFT-LDA spectrum
worsens the comparison with experiments in the present case.
An effect of this type is typical for systems where the  excitonic effects are
large. 
The excitonic binding energies infact, can
range from few meV (like in bulk GaAs and Si) to much larger values (of the 
order of the $eV$) in 
insulating oxides \cite{albrecht1,albrecht2,rohlfing}.
The effects of self-energy and excitonic corrections to the DFT-LDA 
$\epsilon_2$ curve roughly correspond to shifts in opposite directions
\cite{albrecht1,rohlfing}.
Referring to the onset of $\epsilon_2$,
one might obtain a rough estimate of the excitonic binding energy from the 
mismatch between GW corrected onset 
and the experimental one. In the c-BN case, a value of more than 5eV 
would occur. 
This value seems to be too high, even in comparison with the 
strongest exciton binding energies found in systems like oxides,
 where, due to the weak screening, 
the electron and the hole can be bound by an energy  binding of the order of 
1eV \cite{albrecht1}. 
A theoretical evaluation based on Wannier functions for c-BN leads to a 
value of the this order of magnitude \cite{adragna}. 
To fully address this problem from the theoretical point of view,
one should solve the Bethe-Salpeter equation for the optical response 
function including two particle effects \cite{albrecht1,albrecht2,rohlfing}.
On the other hand, one should look for more refined experimental 
mesaurements  of the linear optical properties of c-BN. 
In summary  c-BN and h-BN mixing within the sample, 
self-energy and excitonic effects, 
the presence of impurites, the lack fo pure crystalline samples,
and lattice parameter mismatch, 
can in principle all be sources of the disagreement 
between theoretical results and the available spectra for c-BN.
\section{Conclusions}
Linear optical functions of cubic and hexagonal BN have been studied within 
DFT-LDA.  Calculated energy-loss functions compare
well with experiments and with previous theoretical results, both for
h-BN and c-BN. Discrepancies arise between theoretical
results and experiments in the imaginary part of the dielectric function
for c-BN. Possible explanations to this issue are proposed and evaluated;
lattice constant variations, h-BN contamination in c-BN samples, and 
self-energy effects. On the other hand our DFT-LDA results show a 
reasonable agreement with other theoretical outcomes and with experiments for 
the imaginary part of the dielectric function in the case of h-BN. More
refined measurements and calculations are needed to fully address the mismatch between 
theory and experiment for the cubic case.
It is in our programs to go further in direction of a refinement of the 
calculations.
\section{Aknowledgements}
We thank R. Del Sole, P. Monachesi,G. Adragna, and A. Marini for useful 
discussions. We acknoledge the 
help of G. Pusceddu for generating pseudpotentials used in this work.

%

%--------------------------------tables---------------------------
\begin{table}[]
\centering\begin{tabular}{|r|c|c|c|c|c|}
\hline
    c-BN        &Work1    &Work2    &Theo1        &Theo2        &Exp.  \\
\hline
  $a_0(a.u.)$   &6.754    &6.771    &6.759        &6.833        &6.833  \\
\hline
  $B_o(Mbar)$   &4.01     &3.52     &3.97         &3.70         &3.69-4.65  \\
\hline
\hline
  $h-BN$        &         &         &             &             &      \\   
\hline
\hline
  $a_0(a.u.)$   &4.698    &4.68    &4.50          &4.71         &4.72  \\   
\hline
  $c/a$         &2.608 &2.6068  &2.608        &2.670         &2.664  \\   
\hline
  $B_o(Mbar)$   &2.68     &2.65    &2.61          &3.35         &  \\   
\hline
\end{tabular}
\label{t5}
\caption{Calculated structural properties of zincblende and hexagonal BN
(lattice constant and bulk modulus) compared with other theoretical and 
experimental results. First column: present calculations; second column:
theoretical results by Cappellini et. al.\protect\cite{cappefiore}, 
Theo1: results by  Furthm\"uller et al.\ \protect\cite{furthmueller}, Theo2: 
results by Xu e Ching \ \protect\cite{xu}, and in the last column the 
experimental values from Ref.\protect\cite{cappefiore}.}
\end{table}

\begin{table}[]
\centering\begin{tabular}{|r|c|c|c|c|}
\noindent
%\hline
{c-BN}                       & GW1    & GW1bis & GWf  & Exp          \\
\hline
{$\Gamma_{15'v}-X_{1c}$}     & 7.28    & 6.95   & 6.3  & $6.4\pm 0.5$  \\
\hline
{$\Gamma_{15'v}-\Gamma_{1c}$}& 11.79   & 11.46  & 11.4 & 14.5          \\
\hline
\hline
\end{tabular}
\caption{Excitation energies for c-BN. In the first column, the 
present theoretical values here calculated (GW1), GW1bis is taken from 
 Ref.\protect\cite{cappefiore},
GWf from Ref.\protect  \cite {suhr}. The experimental results
for the indirect gap are from  Ref.\protect \cite {chrenko}, and those for
the direct one are from Ref.\protect \cite {philiz}.} 
\label{tab:2}
\end{table}
\begin{table}[]
\centering\begin{tabular}{|r|c|c|c|c|}
\noindent
%\hline
{h-BN}               & GW1      & GW1bis   & GWf         &Exp         \\
\hline
{$H_{3v}-M_{1c}$}    & 6.39   &  6.04    & 5.4$^a$     & $5.2\pm0.2$ \\
%\hline
%{$M_{3v}-M_{1c}$}    & 6.34  &  6.33    & 7.8         &             \\
\hline
{$H_{3v}-H_{2c}$}    & 6.76  &  6.66    & 6.33$^b$    &        \\
\hline
\hline
\end{tabular}
\caption{Excitation energies for h-BN . The first column contains the
values calculated here (GW1), GW1bis is taken from Ref.\protect\cite{cappefiore},
GWf from Ref.\protect \cite {blase2} and the experimental values from
Ref.\protect \cite {hoffman}. For the meaning of the a and b marks, see 
text.} 
\label{tab:3}
\end{table}

%******************figures

%\begin{figure}
% \epsfxsize=12.0cm
% \epsfbox{riflehex.ps}
% \epsfbox{riflenew.eps}
%\vspace*{2.0cm}
% \caption{Calculated reflectivity of hexagonal BN}
% \label{reflechex}
%  \end{figure}

%***********insert

%**figure
%**

%\end{multicols}

\end{document}